\documentclass[journal]{rmaa}

\usepackage{paralist}

\usepackage{psfrag,color}

\usepackage[latin1]{inputenc}

\SetYear{2008}

\title{Chemical abundances and gas content in disk galaxies:
correlations with the $\lambda$ spin parameter.} 

\author{
  B. Cervantes-Sodi,\altaffilmark{1} 
  X. Hernandez\altaffilmark{1,2}
       }

\altaffiltext{1}{Instituto de Astronom\'\i{}a,
  Universidad Nacional Aut\'onoma de M\'exico, M\'exico.}
\altaffiltext{2}{GEPI, CNRS UMR 8111, Observatoire de Paris, Meudon Cedex, France.
}

\shortauthor{B. Cervantes-Sodi \& X. Hernandez}
\shorttitle{Chemical abundances and gas content vs. $\lambda$}

 \fulladdresses{
 \item B. Cervantes-Sodi: Instituto de Astronom\'\i{}a, Universidad Nacional
  Aut\'onoma de M\'exico, Apdo. Postal 70-264, 04510 M\'exico, D. F., M\'exico
 (bcervant@astroscu.unam.mx).
 \item X. Hernandez: GEPI, CNRS UMR 8111, Observatoire de Paris, 5 Place Jules Janssen, 92195 
Meudon Cedex, France and  Instituto de Astronom\'\i{}a, Universidad Nacional
  Aut\'onoma de M\'exico, Apdo. Postal 70-264, 04510 M\'exico, D. F., M\'exico  (xavier@astroscu.unam.mx)
}

\listofauthors{B. Cervantes-Sodi \& X. Hernandez}
\indexauthor{B. Cervantes-Sodi}
\indexauthor{X. Hernandez.}

\abstract{ 
By using a very simple and general model to describe the dynamics of disk 
galaxies, we estimate the $\lambda$ spin 
parameter for a sample of observed galaxies 
and present a study in which we show that several important physical 
properties are intrinsically related to the dynamics of the systems. 
Although correlations between average metallicity with magnitude or Hubble 
type are evident, we obtain equally 
strong correlations with the spin parameter, where galaxies with low 
$\lambda 
$ values present higher abundances 
and galaxies with high $\lambda$ values are poor in 
metals. Also, the gas content of the galaxies correlates with $\lambda$, 
with 
high $\lambda$ systems showing 
higher gas mass fractions than low $\lambda$ galaxies, highlighting the 
important role 
this parameter plays in the structure of disk galaxies and the proposal of 
$\lambda$ as a robust and objective physical measure of galactic 
morphology.

}

\resumen{

Usando un modelo simple y general para describir la din\'amica de las 
galaxias de disco, estimamos el 
parametro de esp\'\i{}n $\lambda$ para una muestra de galaxias observadas 
y 
presentamos un estudio que muestra 
como varias propiedades f\'\i{}sicas se 
encuentran relacionadas intr\'\i{}nsecamente con la din\'amica de los 
sistemas. A\'un cuando las correlaciones entre metalicidad promedio y 
magnitud o tipo de Hubble son evidentes, nosotros obtuvimos correlaciones 
igualmente fuertes con el  esp\'\i{}n, en el sentido que galaxias con 
valores 
peque\~{n}os de $\lambda$ presentan abundancias mas altas, mientras que 
galaxias con $\lambda$ grande son pobres en metales. Tambi\'en el 
contenido 
de gas en las galaxias correlaciona con $\lambda$, 
en donde sistemas con valores altos de $\lambda$ presentan mayores 
fracciones 
de masa de gas que galaxias con valores peque\~{n}os de $\lambda$, 
destacando el importante papel que juega este par\'ametro en la estructura 
de 
las galaxias de disco y la propuesta 
de su uso como una medida del tipo morfol\'ogico.
}

\addkeyword{galaxies: abundances}
\addkeyword{galaxies: general}
\addkeyword{galaxies: structure}
\addkeyword{galaxies: fundamental parameters}

\begin{document}
\maketitle

\section{Introduction}
\label{sec:intro}

One recurrent issue in the study of galaxies, is their classification, most frequently 
treated through the Hubble type (Hubble 1926, 1936). The extended use of this  
classification scheme owes it success to 
the ample and broad range of physical characteristics involved in its construction, 
such that all show good monotonic correlations with it. Just to mention some of them, 
total magnitudes decrease towards later
types (e.g. Ellis et al. 2005), and colours become bluer from early to late types 
(e.g. Roberts \& Haynes 1994), the blue magnitudes and bulge to disk ratios decrease 
(e.g. Pahre et al. 2004) and disks become thinner (e.g. Kregel et al. 2002) in going to late types.

All this clear correlations point towards the notion that the sequence of morphologies indicate an underlying
 sequence of values for one or more physical parameters, which ultimately determine the type 
of the galaxies. The Hubble scheme of
 classification owes its success precisely to this fact; it orders galaxies by properties that reflect
 essential physics, but not without some shortcomings. The classification of galaxies using this scheme
 is usually done by visual inspection and requires skill and experience, which not only makes it a subjective scheme,
it also becomes impractical when dealing with large samples of galaxies containing hundreds or thousands of galaxies,
 as is the case of the Sloan Digital Sky Survey (SDSS) or the 2dF Galaxy Redshift Survey. Naim et al. (1995)
 and Lahav et al. (1995), studied the agreement between expert morphologists classifying different samples
 of galaxies and both collaborations reported a non-negligible scatter between observers, 
making the subjective character of the Hubble scheme evident.

Another important problem with the Hubble scheme, is that it is based on a qualitative analysis of
 observable features, which makes it difficult to relate type to definitive quantitative  physical aspects of a
 galactic system. That the scheme is ultimately qualitative sheds serious doubts on the validity of
any statistical analysis performed on galactic populations based on galactic type.

The study of high-z galaxies whose light has been redshifted into other bands, shows a wavelength
 dependence of morphology for different Hubble types. In general, it is found (Kuchinski et al. 2000)
 that the dominance of young stars in the far-ultraviolet produces the patchy appearance of a morphological
 type later than inferred from optical images. Apart from the observational band dependence, the low signal-to-noise 
levels in faint images of distant galaxies, projection effects and low resolution, can strongly affect
 the morphological type assigned to a given galaxy (e.g. Abraham et al. 1994). This makes the Hubble
scheme relative to the observation band, the redshift and the orientation available for a given galaxy,
as the spiral arms vanish for edge on disks.

From the theoretical study of galactic structure and formation, it generally emerges that the principal 
physical origin of galactic type morphology, lies in the choice of mass and angular momentum.
Some examples of the above are:  Fall \& Efstathiou (1980), Firmani et al. (1996), Dalcanton, Spregel \& Summers (1997), 
Mo, Mao \& White (1998), and van den Bosch (1998). Theoretically, the observed spread in many physical 
features of the galaxies can be understood as arising from a spread in the halo angular momentum once 
the mass has been fixed, coupled with a spread in the formation redshifts, this last, once
a particular formation model has been assumed.
For example, the bulge 
to disk ratio (van den Bosch 1998), the thickness of the disk (Kregel, van der Kruit \& Freeman 2005), 
the surface density (Dalcanton, Spregel \& Summers 1997), systematic variations in the slope of rotation
 curves (Flores et al. 1993), scalelength (Hernandez \& Gilmore 1998, Jimenez et al. 1998), chemical
 abundances and color gradients (Prantzos \& Boissier 2000), and gas content, abundance gradients
 (Churches, Nelson \& Edmunds 2001, Boissier et al 2001) and the spread in the Tully-Fisher relation, once the slope is
 determined by the total galactic mass (Koda, Sofue \& Wada 2000), to mention but a few. Hence, the proposal 
of the angular momentum as the principal physical parameter responsible for the Hubble 
sequence (Silk \& Wyse 1993, van den Bosch 1998, Zhang \& Wyse 2000, Firmani \& Avila-Reese 2003)
is not surprising. It has often been suggested that the angular momentum of a galaxy should be
an adequate proxy for the Hubble sequence, or rather, that it is precisely this parameter
the one for which the Hubble sequence has been serving as a proxy off.

In Hernandez \& Cervantes-Sodi (2006), henceforth Paper I, using a sample of observed galaxies, we presented 
a series of correlations confirming the dependence on angular momentum of several key structural parameters
 such as disk to bulge ratio, thickness of the disk and rotational velocity, and in Hernandez et al. (2007, Paper II),
 we matched Hubble type with the angular momentum parameter using a color versus color gradient criteria
 (Park \& Choi 2005), where the segregation by angular momentum coincides with the segregation by Hubble type,
as visually assigned for a large sample of spiral galaxies from the SDSS.

In the present work, we give a general review to introduce the use of the angular momentum parameter as
a physical classification criterion for spiral galaxies, which we reinforce by presenting additional independent 
correlations recently found. In Section 2 we present the underlying ideas behind proposing physical
correlations between structural galactic parameters and angular momentum,
 and an account of previous results, in Section 3 we explore possible scalings between chemical abundances, 
gas content and angular momentum, which we test in Section 4. 
Section 5 presents a discussion of our results and general conclusions.

\section{Theoretical framework}
\label{sec:framework}

A generally accepted picture for galaxy formation is the $\Lambda$CDM model, where overdense regions 
of the expanding Universe at high redshift, accrete baryonic material through their gravitational potential to become
 galaxies. Their principal global characteristics, according to theoretical studies, are determined by their mass and
angular momentum. The angular momentum is commonly characterized by the dimensionless angular
 momentum parameter (Peebles 1969):

\begin{equation}
\label{Lamdef}
\lambda = \frac{L \mid E \mid^{1/2}}{G M^{5/2}}
\end{equation}

where $E$, $M$ and $L$ are the total energy, mass and angular momentum of the configuration, respectively.
 In Paper I we derived a simple estimate of $\lambda$ for disk galaxies in terms of observational parameters,
 and showed some clear correlations between this parameter and structural parameters, such as the disk 
to bulge ratio, the scaleheight and the colour. Here we briefly recall the main ingredients of the simple
 model. The model considers only two components, a disk for the baryonic component with an exponential
 surface mass density $\Sigma(r)$;

\begin{equation}
\label{Expprof}
\Sigma(r)=\Sigma_{0} e^{-r/R_{d}},
\end{equation}

\begin{figure*}[htb]
  \makebox[0pt][l]{\textbf{a}}%
  \hspace*{\columnwidth}\hspace*{\columnsep}%
  \textbf{b}\\[-0.7\baselineskip]
  \parbox[t]{\textwidth}{%
     \vspace{0pt}
     \includegraphics[width=\columnwidth,height=6cm]{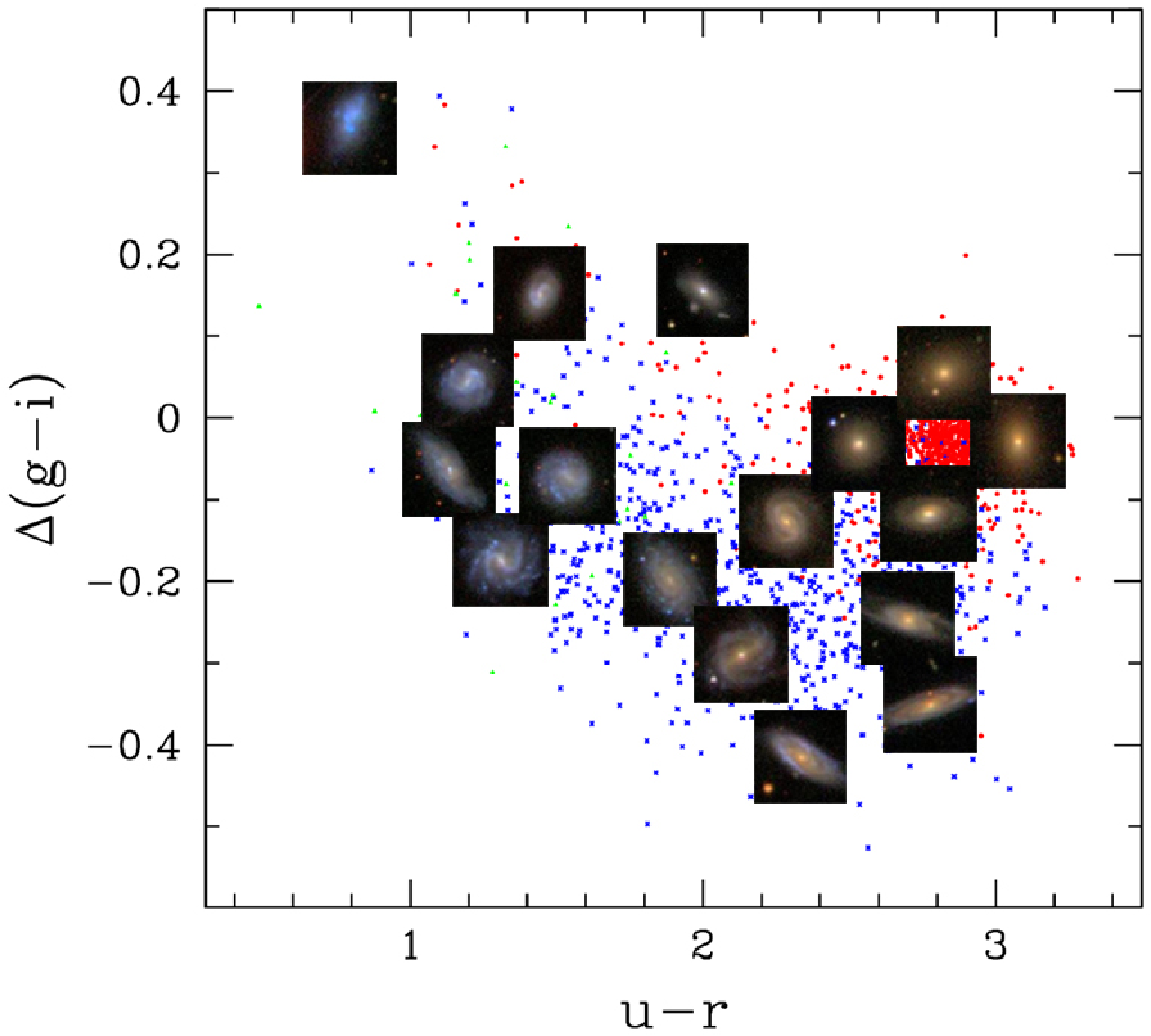}%
     \hfill%
     \includegraphics[width=\columnwidth,height=6.25cm]{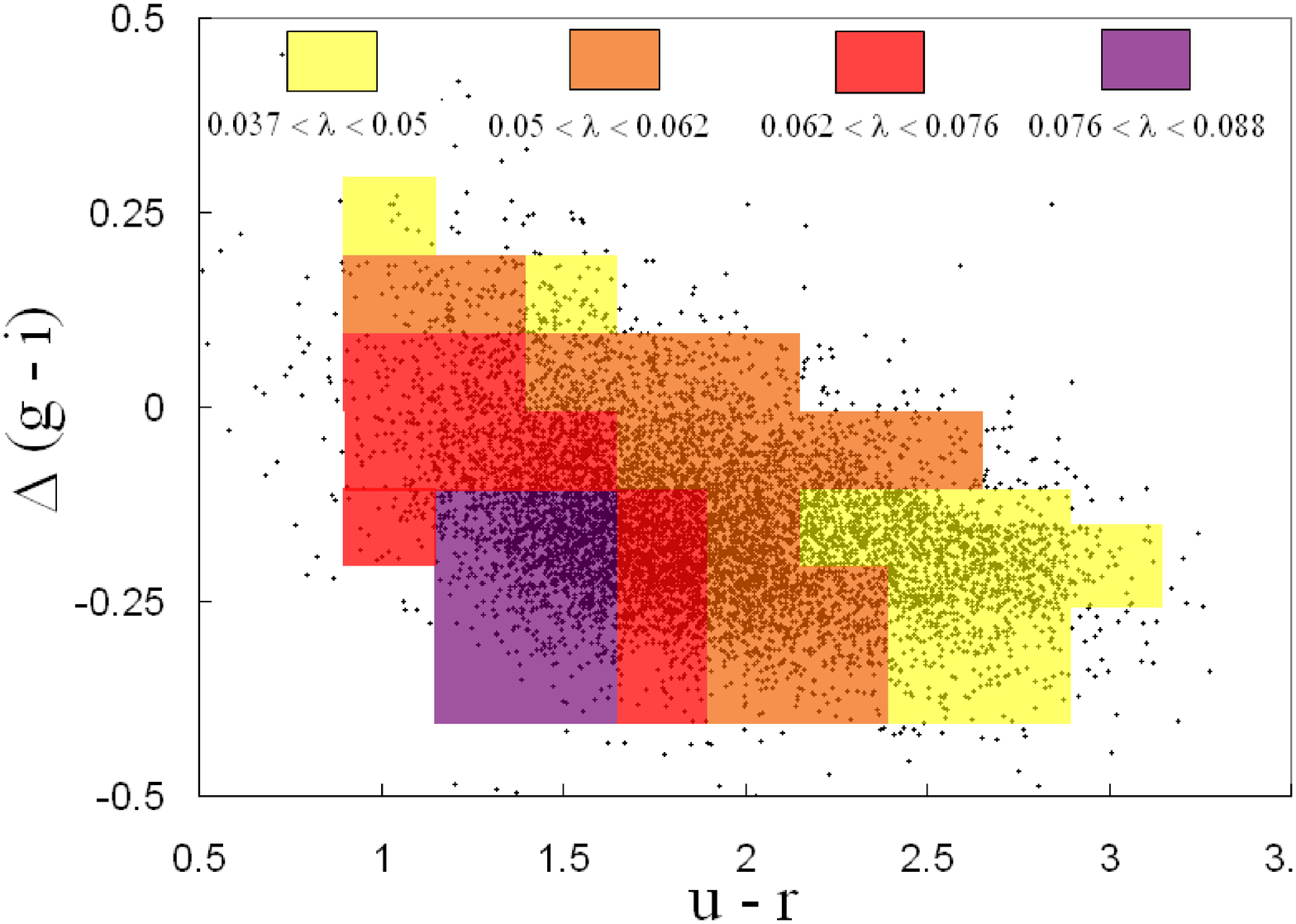}
     }
  \caption{(\textit{a})Sample of galaxies from Park \& Choi (2005) in a color, color-gradient plane, with photos of galaxies representative of the morphological type found in each region. 
    (\textit{b})Spiral galaxies from Paper II on a color, color-gradient plane, the shading shows the average $\lambda$ value in each shaded region, calculated using equation (5).  }
  \label{fig:types}
\end{figure*}

where $r$ is a radial coordinate and $\Sigma_{0}$ and $R_{d}$ are two constants which are allowed to
vary from galaxy to galaxy, and a dark matter halo having a singular isothermal density profile $\rho(r)$,
 responsible for establishing a rigorously flat rotation curve $V_{d}$ throughout the entire galaxy;

\begin{equation}
\label{RhoHalo}
\rho(r)={{1}\over{4 \pi G}}  \left( {{V_{d}}\over{r}} \right)^{2}. 
\end{equation}

In this model we are further assuming: (1) that the specific angular momentum of the disk and halo are equal
 e.g. Fall \& Efstathiou (1980), Mo et al. (1998), Zavala et al. (2007); (2) the total energy is dominated
 by that of the halo which is a virialized gravitational structure; (3) the disk mass is a constant
 fraction of the halo mass; $F=M_{d}/M_{H}$ (Brooks et al. 2007, Crain et al. 2007). These assumptions
 allow us to express $\lambda$ as

\begin{equation}
\label{Lamhalo}
\lambda=\frac{2^{1/2} V_{d}^{2} R_{d}}{G M_{H}}.
\end{equation}

Finally, we introduce a baryonic Tully-Fisher relation: $M_{d}=A_{TF} V_{d}^{3.5}$, in
consistency with Gurovich et al. (2004) and Kregel et al. (2005). We see the total
mass fixing the bulk dynamical structure of the galaxy, with internal distributions
being then determined by the choice of $\lambda$, e.g. Fall \& Efstathiou (1980)
and references thereof. Taking the Milky
Way as a representative example, we evaluate $F$ and $A_{TF}$ to obtain

\begin{equation}
\label{LamObs}
\lambda=21.8 \frac{R_{d}/kpc}{(V_{d}/km s^{-1})^{3/2}}.
\end{equation}

The accuracy in our estimation of $\lambda$ is proved in Cervantes-Sodi et al. (2008), comparing the estimation using
equation (5) with the values arising from numerical simulations of six distinct groups, where the actual
value of $\lambda$ is known, as it is one of the parameters of the simulated galaxies, and where we can also
estimate it through equation (5), as baryonic disk scale lengths and disk rotation velocities are part of the output.
This shows a one-to-one correlation, with very small dispersion leading to typical errors $<$ 30\%.
This level of accuracy is encouraging, when comparing what is little more than an order of magnitude estimate
against detailed physical modeling spanning a wide range of masses and $\lambda$ values, including
disk self-gravity, the presence of galactic bulges, detailed dark halo structure and often complex formation
histories. The only ambiguity which remains in assigning a value of $\lambda$ to an observed galaxy
through equation (5), is in the choice of restframe band in which the disk scalelength is to be measured.
This is clearly to be chosen in a way which maximally reflects the underlying mass distribution, and not
the recent star formation, highly sensitive to a small number of young stars. Therefore, the restframe
band in which $R_{d}$ is to be measured if equation (5) is to be used optimally, is a red or infrared one,
see Paper I.

Having an estimate for the spin parameter, we applied it to a large sample of disk galaxies taken from 
the SDSS. Park \& Choi (2005), showed a criteria using the colour versus 
colour gradient plane, to separate galaxies by galactic type, from early to late type (see figure 1, left panel).
We see the ellipticals appearing in a very compact region, with red colours and minimal colour gradients.
In figure 1 right panel, are plotted the 7,753 galaxies from the SDSS sample of Paper II, in the same color versus colour gradient
 plane, where the shading gives the average values of $\lambda$ within each shaded square, calculated
through equation (5). We can see that
 high $\lambda$ disk galaxies are located in the lower left area of the plane, occupied by
later types in figure 1 left panel, whilst low $\lambda$ disk
 galaxies populate the right upper zone, precisely where earlier types are evident in figure 1 left panel. 
Comparing with the spread of galactic types, the match of late
 spirals with high $\lambda$ values and early spirals with low $\lambda$ values is validated.

With equation (5), in Paper I we showed several correlations between $\lambda$ and physical features of
disk galaxies such as: the disk to bulge ratio, one of the main morphological parameters that sets the
classification of galaxies in the revised Hubble diagram (van den Bosch, 1998), which decreases when
$\lambda$ increases, while the thickness of the disk diminishes and the colour becomes bluer. 
A clear trend for mean $\lambda$ values increasing in going towards later Hubble types, as visually assigned,
was also established. In the 
following section we present the underlying ideas behind expecting additional correlations involving gas fractions and chemical abundances,  with the aim of reinforcing 
the use of $\lambda$ as an objective and quantitative classification tool.

\section{Expected correlations between $\lambda$ and chemical Abundances and gas content}
\label{sec:chemestry}

The gas mass and chemical composition in disk galaxies span wide ranges, early-type galaxies presenting 
systematically lower gas mass fractions and higher metallicity abundances (Vila-Costas \& Edmunds 1992, 
Oey \& Kennicutt 1993, Roberts \& Haynes 1994 and Zaritsky et al 1994). As pointed out by Kennicutt et al. (1993),
it is tempting to associate the higher abundances with the early morphological types, but differences in other parameters
such as galaxy luminosity or rotational velocity may also be important. Over the last twenty years there have
been several studies exploring the connection between metallicity and other physical parameters of galaxies; 
luminosity, rotational velocity and mass, among others.

The correlation between metallicity and luminosity, in the sense that the higher the
luminosity of the system, the higher the metal abundance, is now well established e.g.
Garnett \& Shields (1987), Hunter \& Hoffman (1999),
Pilyugin (2001) and Melbourne \& Salzer (2002). Thinking of the luminosity as a tracer for the mass of the galaxy, 
and given the Tully-Fisher relation, we find equally a metallicity rotation velocity relation. Compared with the luminosity
metallicity relation, the above shows no 
qualitative difference (Garnett 1998), only a slightly better correlation, presumably due to differences
among mass to light ratios. 
The mechanism that might be responsible for the origin of such correlations
is, at present, open to debate, but the most common hypothesis is the loss of heavy elements through galactic 
winds (e.g., Franx \& Illingworth 1990, Kauffmann 1996, Pilyugin, V\'\i{}lchez \& Contini 2004) in low 
mass galaxies, while higher mass galaxies present deeper gravitational potentials and lower heavy elements 
losses. The correlation with Hubble type however, is more difficult to explain given the fact that this last is not a 
physical parameter as such.

If besides the mass, there is another physical parameter which determines abundances in galaxies, it should 
correlate with the Hubble type. It is reasonable to think that the angular momentum parameter could play an 
important roll. In order to see how a causal relation between
$\lambda$ and both, the characteristic abundances and gas content of 
galaxies might arise, one can start from the star formation rate (SFR), assuming a generic Schmidt law in terms of SFR 
surface density and gas surface density:

\begin{equation}
\label{SFR}
\Sigma_{SFR} \propto \Sigma^n.
\end{equation}

We define the star formation efficiency $\epsilon$ evaluated at $R_{d}$ as:

\begin{equation}
\label{SFE}
\epsilon = \frac{\Sigma_{SFR}(R_{d})} {\Sigma(R_{d})} \propto \Sigma_{0}^{n-1},
\end{equation}

and using equation (2), we can write $\Sigma_{0}$ in terms of the mass of the disk and the scalelength,
from where the dependence in $R_{d}$ can be replaced by a dependence in $\lambda$ at fixed mass, resulting in 

\begin{equation}
\label{epsilon}
\epsilon \propto \lambda^{2(1-n)}.
\end{equation}

Both empirically (Kennicutt 1998, Wong \& Blitz 2002) and  theoretically (Tutukov 2006) inferred values
for $n$ fall in the range $1-2$, being 
$1.4$ a standard value,
in this case, we have $\epsilon \propto \lambda^{-0.8}$, in agreement with previous results from theoretical 
studies (Boissier, Boselli, Prantzos \& Gavazzi 2001). For any value of $n$ in the cited range 
(excluding one of the extremes), the exponent
in equation (8) is negative, so systems with large $\lambda$ values will show typically low star formation
efficiencies, while galaxies with small $\lambda$ may present a very efficient star formation process. In 
other words, extended galaxies, with large scalelengths for their total mass, present lower
star formation efficiencies than more compact ones, those with small scalelength at a given mass.
The scaling corresponding to equation(8), for $\epsilon$ as a function of mass at fixed $\lambda$
yields $\epsilon \propto M^{(n-1)/7}$. For standard values of n, we see the expected positive correlation of 
$\epsilon$ vs. mass, opposite to what happens vs. $\lambda$, leading to an increasingly later type appearance
as one goes to larger masses, at fixed $\lambda$.

Studying
the chemical and spectrophotometric evolution of galactic disks, using semi-analytic models of galactic evolution,
Boissier \& Prantzos (2000) reached similar conclusions. In their models, the shortest characteristic time-scales corresponds
to the more massive and compact disks with smaller $\lambda$ values, favoring rapid early star formation and the
existence of relatively old stellar populations today.

From equation (8), the most efficient galaxies at forming stars and therefore metals, are the compact ones, 
with low $\lambda$ values, being also the most efficient at turning their gas into stars. If we assume that
 all galaxies start forming stars at the same time and that the amount of gas in each of them is proportional
 to their gravitational potential, extended galaxies with high $\lambda$ should appear today to be systems with large 
amounts of gas and poor in metals, while compact galaxies, with low $\lambda$ and very efficient at forming
stars at the onset, will now be systems poor in gas with large metal abundances. Prantzos \& Boissier (2000) reported this same result
arising from their semi-analytic models, where the most compact disks with low $\lambda$ values presented the higher
abundances, being this type of galaxies the most efficient at forming stars and metals, consuming their gas and showing
low gas mass fractions. The above applies only to large and medium disk galaxies, the clear mass metallicity relation in spheroidal systems, from dSphs to elliptical galaxies, systems where the angular momentum plays only a very minor role, most probably arises from the differential efficiencies of galactic winds. We see how the two parameter
nature of galactic systems reduces to a one parameter space when one goes from the spirals to the ellipticals,
where the angular momentum becomes close to negligible and ceases to determine the internal structure of the galaxy
to any substantial level.

\begin{figure}[htb]
\begin{tabular}{l}
\includegraphics[width=\columnwidth]{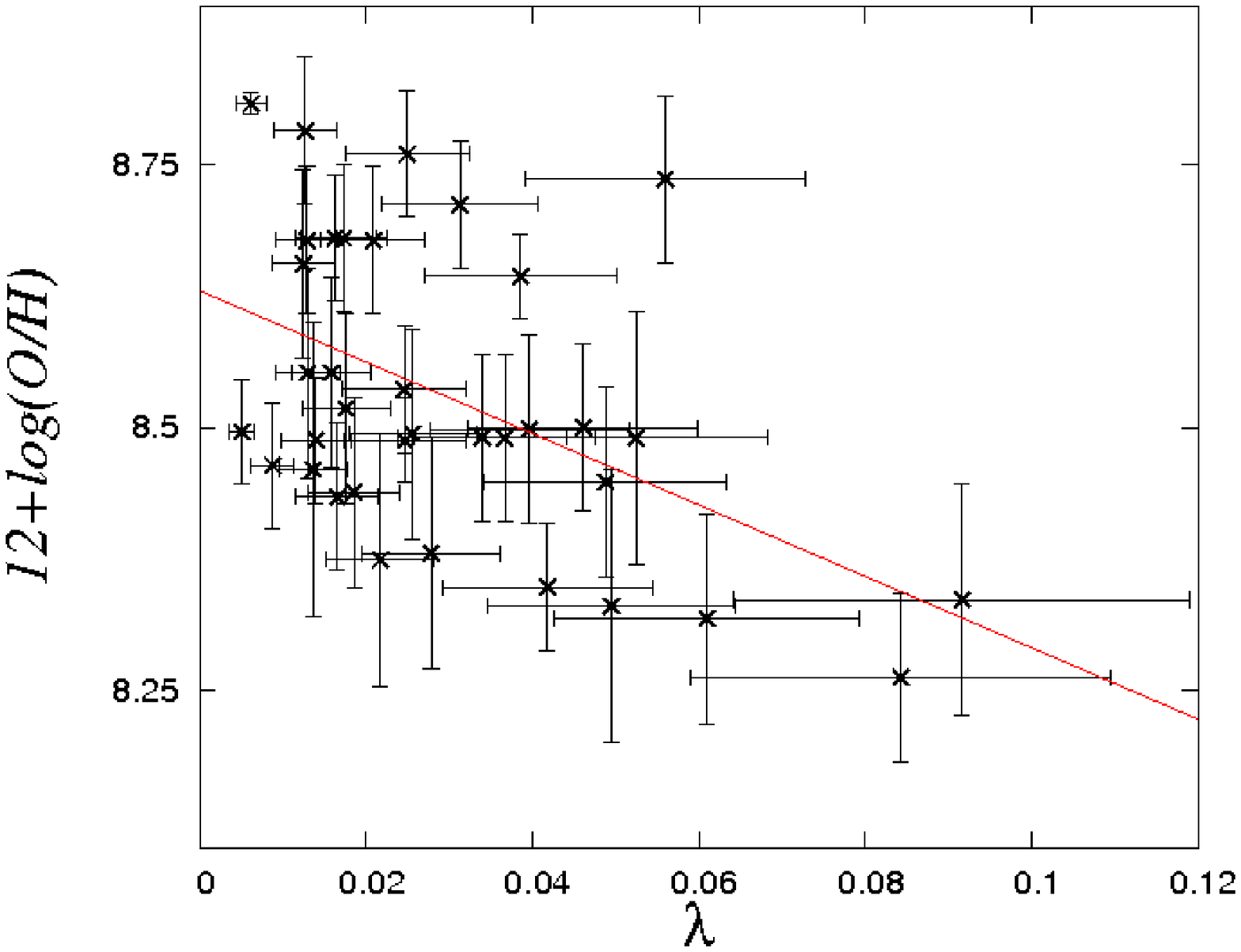}\\
\includegraphics[width=\columnwidth]{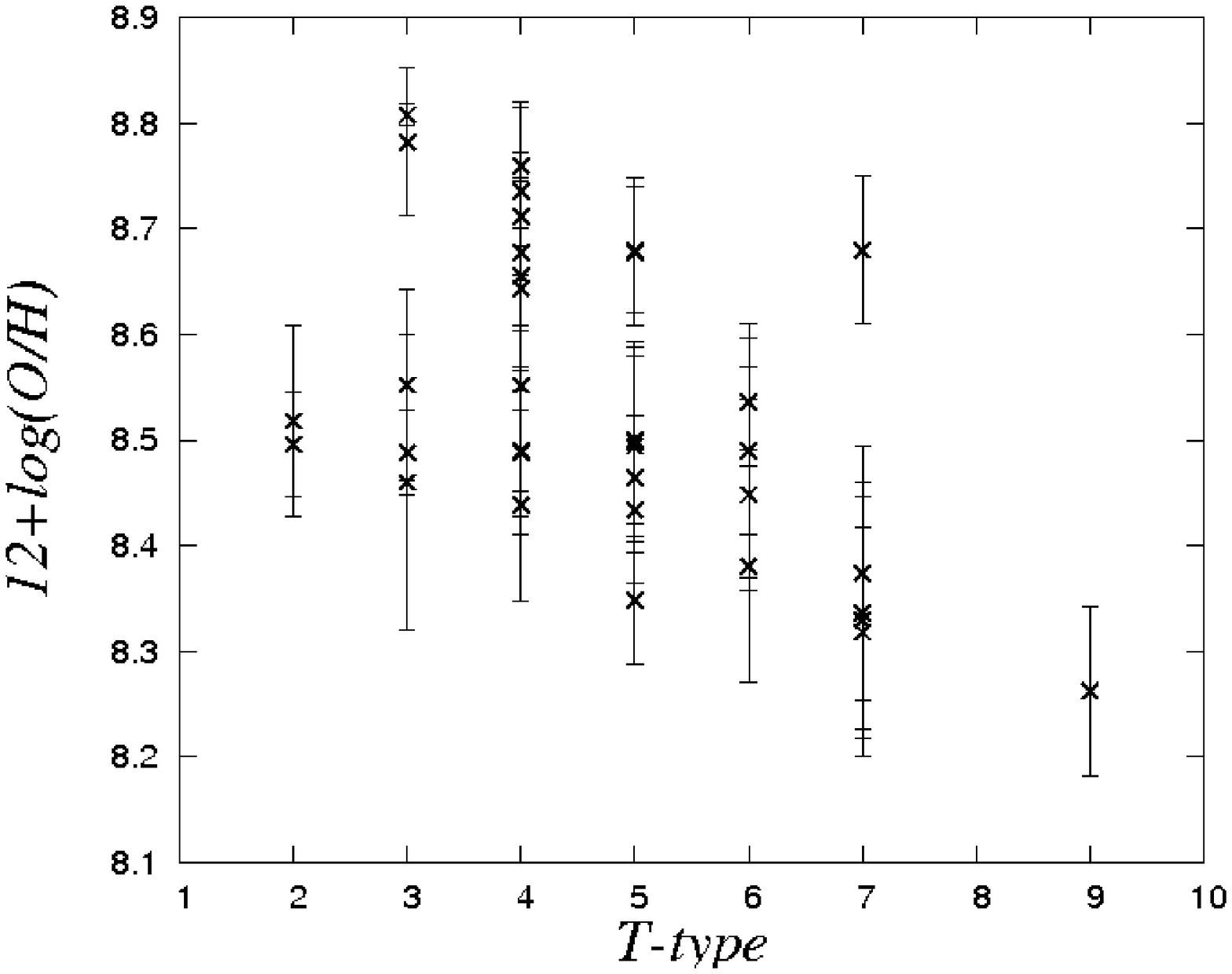}
\end{tabular}
\label{histograms}
\caption[ ]{Characteristic oxygen abundance as a function of $\lambda$ (top panel) and type (bottom panel) 
for the 36 galaxies in the PVC sample for which disk scalelenghts in red bands are available. The solid line is the best linear fit to the data.}
\end{figure}

\section{Observational Data Comparisons}
\label{sec:EData}

\begin{figure*}[htb]
  \makebox[0pt][l]{\textbf{a}}%
  \hspace*{\columnwidth}\hspace*{\columnsep}%
  \textbf{b}\\[-0.7\baselineskip]
  \parbox[t]{\textwidth}{%
     \vspace{0pt}
     \includegraphics[width=\columnwidth]{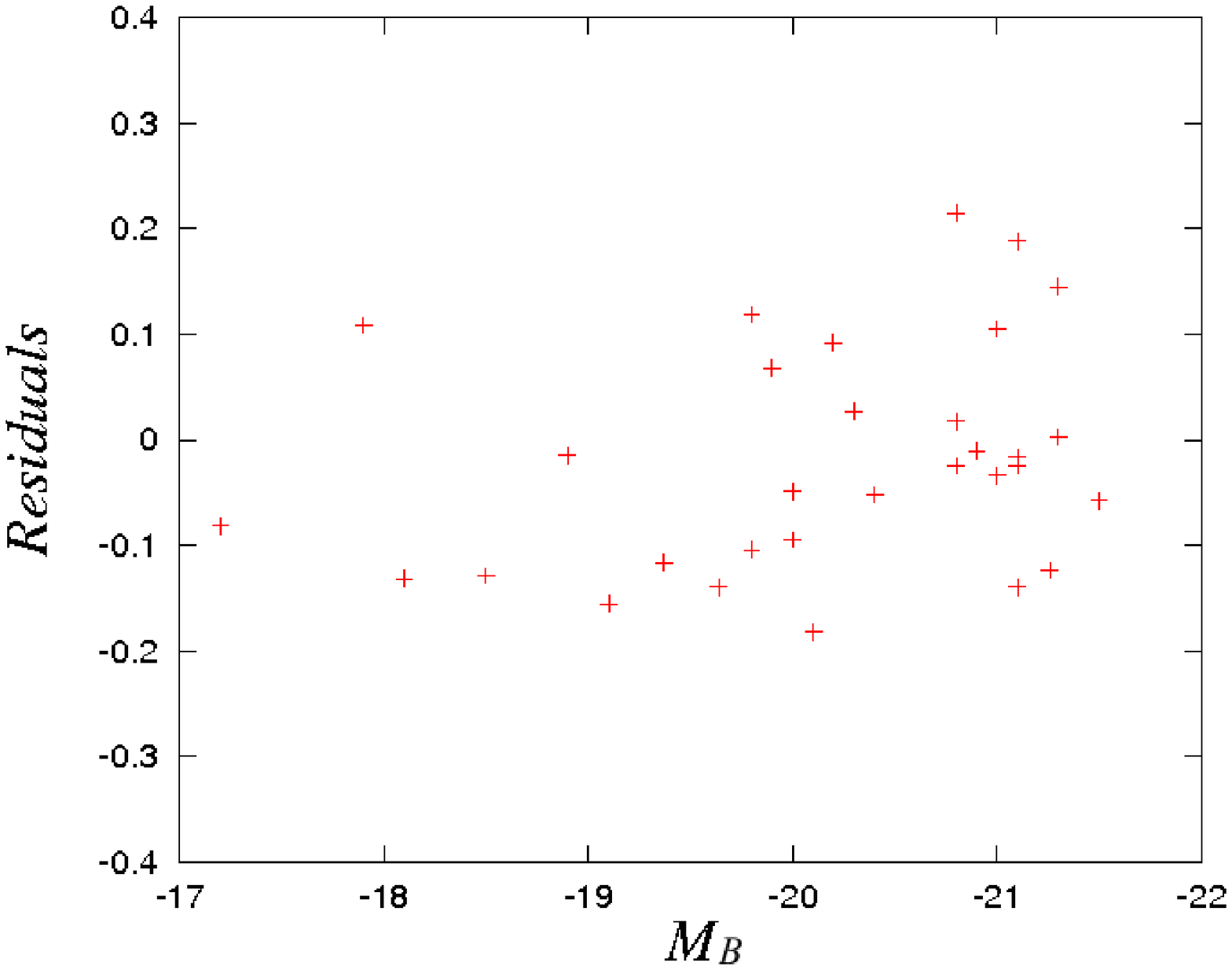}%
     \hfill%
     \includegraphics[width=\columnwidth]{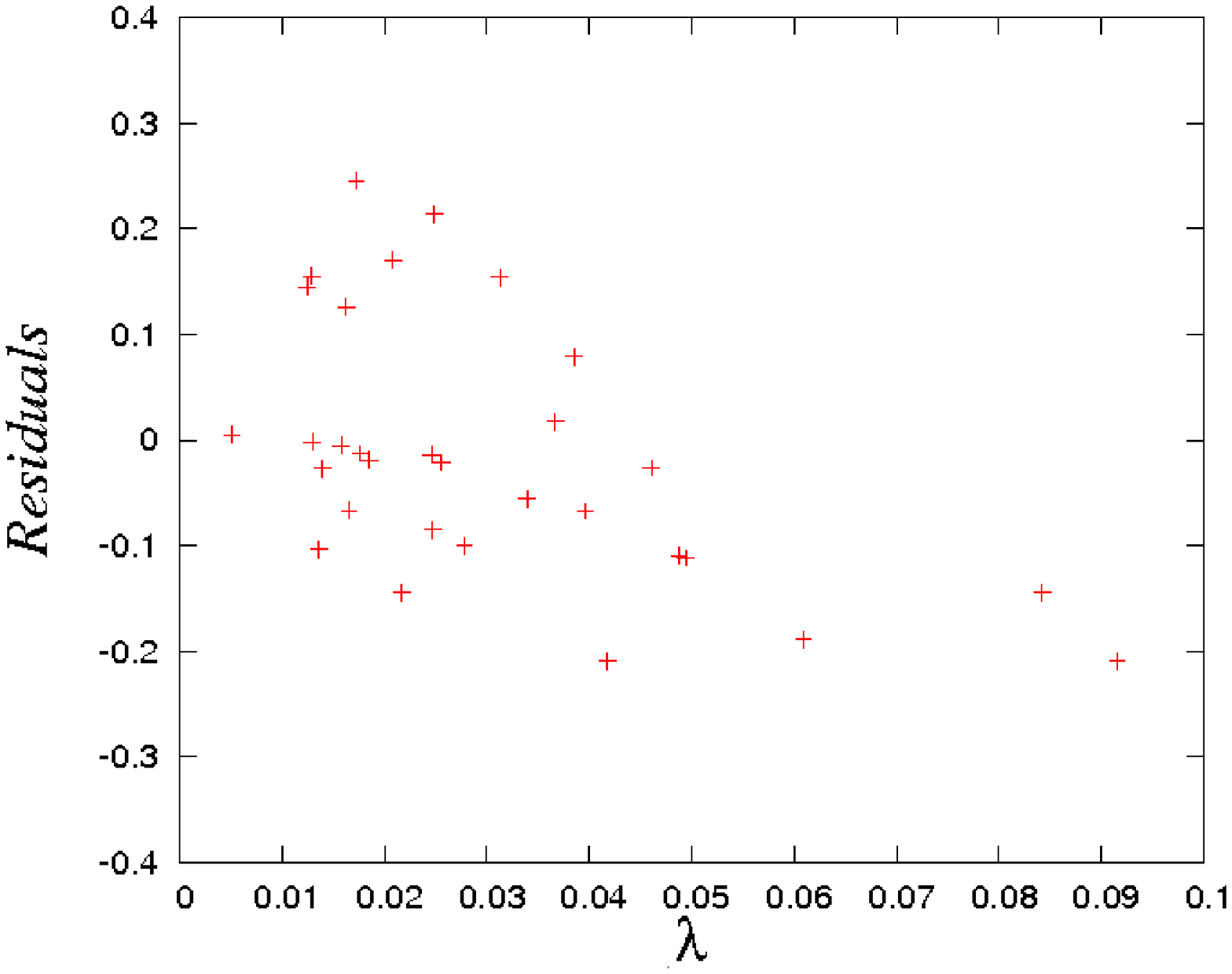}
     }
  \caption{The residuals to correlations. (\textit{a}) Residuals to the $O/H - \lambda$ relationship,
 plotted against $M_{B}$.
    (\textit{b}) Residuals to the $O/H - M_{B} $ relationship, plotted against $\lambda$.  }
  \label{fig:residuals}
\end{figure*}

The observational data used to check our expected correlations were compiled by
Pilyugin, V\'\i{}lchez \& Contini (2004, PVC). The compilation comprises more than 1000 published spectra of 
HII regions in 54 spiral galaxies. The oxygen and nitrogen abundances were obtained using the P-method
described in Pilyugin (2003).

For the present work, we used the exponential scalelengths measured by
Garnett (2002), Zaritsky, Kennicutt \& Huchra (1994, ZKH) and
Elmegreen \& Elmegreen (1984), for 36 of the 54 galaxies in the PVC sample, preferentially
using reported values in red bands in order to account for the mass distribution of the galaxy. Using 
the rotation velocity reported in PVC we calculate $\lambda$ for each galaxy in the sample using equation (5). 
As described in the 
previous section, we expect galaxies with low $\lambda$ values to present very efficient early star formation processes,
capable of enriching their interstellar media with metals and hence, presenting today higher oxygen abundances than the
more extended galaxies with higher $\lambda$, where we would expect lower oxygen abundances.
To check this, we employed the characteristic oxygen of the galaxies, defined as the oxygen abundance in
the disk at $r=0.4R_{25}$, following Zaritsky et al. (1994). In figure 2 we show the characteristic oxygen
of the sample as a function of $\lambda$ (top panel) and type (bottom panel), the straight line being the best 
linear fit to the data, described by the equation:

\begin{equation}
\label{o/hvsl}
12+log(O/H)=-3.401(\pm 0.037) \lambda + 8.667(\pm 1.03),
\end{equation}

which gives also the $1 - \sigma$ uncertainties in the parameters of the fit, given our $30\%$ uncertainties in the
estimates of individual $\lambda$s.
The figure shows a clear correlation between the characteristic oxygen and $\lambda$, with a correlation
coefficient of $-0.493$. Despite the large error bars and high dispersion, large empty regions to the upper right and lower left are evident. The above correlation coefficient is comparable to that obtained against rotation velocity of $0.608$,
or against luminosity of $0.469$, for the correlations against characteristic oxygen. 
The average value of the dispersion in characteristic oxygen for the sample used in figure 2 is of
0.136 dex. Although the spread seen in this figure could be intrinsic, it is also consistent with the 
internal errors in the estimates of $\lambda$, in this sense, the metallicity $\lambda$ relation could in 
fact be much tighter than what the figure shows. The good correlation against rotation velocity mentioned
above, given the Tully-Fisher relation, highlights again the equally important role of mass in galactic structure
and evolution, as often recognised.

The trend with $\lambda$ is clearer than the trend with type, the upper panel of figure 2 gives a metallicity
vs. $\lambda$ relation, having the advantage of the entirely objective nature of its calculation 
and of it being quantitative, hence useful for meaningful 
statistics or to establish mathematical relations with other physical parameters such as equation (9). It must
be taken into account that assigned type of each galaxy is entirely subjective and not free of errors due to lack of
information, in this sense we could assign error bars to the data plotted in figure 2, spanning up to three types around
the reported value.

From figure 2, it is clear that the correlation of the metallicity of the galaxies with $\lambda$ is as strong as
that of metallicity against both the rotational velocity and the luminosity or, 
given the Tully-Fisher relation, the mass of the system.
From the data we see that the studied properties of the galaxies are largely determined by 
fixing the mass and $\lambda$, but until 
now, we are not sure if the dependence with $\lambda$ is not only a reflection of the mass metallicity relation. 
In order to separate the 
respective dependencies, we can plot the residuals of the relation of the metallicity with one of the variables
and test if they correlate with the other variable. To trace the mass of the galaxies we chose the absolute magnitude,
because to calculate $\lambda$ using equation (5) we employed $V_{d}$, in this form the measurement will be 
completely independent. In figure~\ref{fig:residuals} are plotted the residuals: (a) to the $O/H - \lambda$ 
relationship, plotted against $M_{B}$ and (b) to the $O/H - M_{B}$ relationship, plotted against $\lambda$.
We can see a correlation in both cases which indicates a systematic variation due to both variables, the mass
and $\lambda$, as we expected. We hence see that the metallicity is a function of both mass and $\lambda$.
As often found in theoretical studies of galactic formation and evolution, we see observable
properties of galaxies appearing as two parameter families of solutions, determined by the choice of 
mass and $\lambda$.

\begin{figure}[htb]
  \includegraphics[width=\columnwidth]{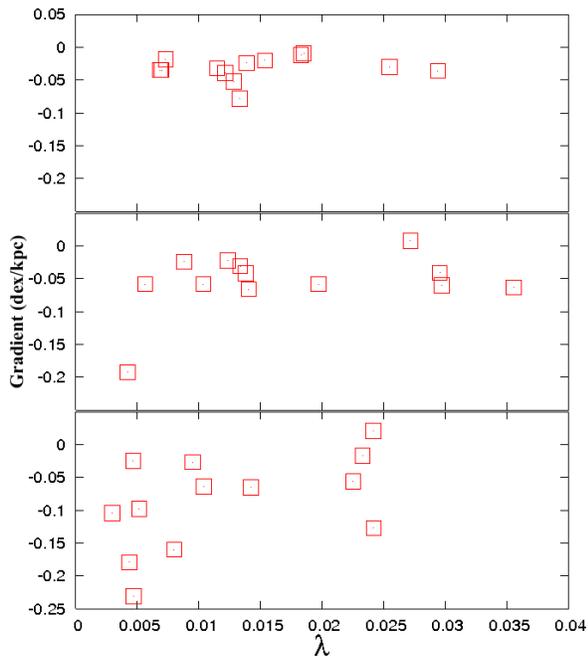}
\caption{Abundance gradients for the 38 galaxies from the ZKH sample, as a function of $\lambda$. From top 
to bottom, subsets of the sample splited by decreasing luminosity}
  \label{fig:grad}
\end{figure}

A similar test was done by ZKH using the
Hubble type and the velocity as a measure of the mass, and they found no clear trends, 
perhaps because the Hubble type is not an objective, quantitative physical parameter, while $\lambda$ satisfies these conditions.

Results from several surveys of galactic abundances show clearly that most disk galaxies present negative 
abundance gradients similar to what is observed in our own galaxy (Henry \& Worthey 1999). While characteristic 
abundances increase with galactic mass, no correlation has being found between metallicity gradients, normalized
to the respective galaxy's isophotal radius, and mass. If instead of using the gradient normalized to the 
isophotal radius, the gradient is measured in $dex$ $kpc^{-1}$, a correlation appears such that more luminous galaxies 
have flatter slopes (Garnett 1998). With this in mind, we search for a dependence of the abundance gradients
with $\lambda$. The abundance gradients of the galaxies in the PVC sample are given in $dex/R_{25}$ and we
find no clear trend with $\lambda$, as predicted in theoretical studies (see below), but trends between $\lambda$ and 
abundance gradients in $dex$ $kpc^{-1}$ have been predicted. Theoretical studies by Churches, Nelson \& Edmunds 
(2001) show that both
the gaseous and stellar abundance gradients in $dex$ $kpc^{-1}$ are functions of $\lambda$ in the sense that 
increasing $\lambda$ generally produces shallower gradients, and Prantzos \& Boissier (2000) predicted that this
effect should be more marked in low-mass disks. 

To study this relation we employed a sample of 38 galaxies from ZKH,
where the oxygen abundance gradients are given in $dex$ $kpc^{-1}$. Using the reported disk scale length and 
the rotation velocity for each galaxy we obtain $\lambda$, again using equation (5). A plot of the abundance
gradient as a function of $\lambda$ is presented in figure~\ref{fig:grad}, where the sample has being split 
into 3 slices, through a luminosity ranking, each one containing one third of the total sample. The top panel presents the galaxies
with the higher luminosities showing almost a constant value and very low dispersion, the middle panel again shows
low dispersion and in general low gradients, but as we move to the smaller galaxies (bottom panel), the dispersion 
increases, showing deeper gradients for galaxies with low $\lambda$, confirming the theoretical predictions of 
Churches, Nelson \& Edmunds (2001) and Prantzos \& Boissier (2000), where in low-mass galaxies, the increase
of $\lambda$ produces shallower gradients. Again, the two parameter nature of galaxies becoming apparent
in the two fold dependence of abundance gradients against both mass and $\lambda$, being predominant, in this case, the roll played by the mass.

An extreme case would be that of the Low Surface Brightness galaxies,
having very extended disks and large $\lambda$ values (Hernandez \& Gilmore 1998) which then should exhibit 
small abundance gradients or none at all (Jimenez, et al. 1998). Observations of de Blok \& van der Hulst (1998), 
shows exactly this behavior, explained as a result of a constant evolution rate over the entire disk. 

The high star formation efficiency of low spin galaxies should be reflected in the gas content of such systems. 
To measure this effect, we employed the gas fraction $\mu$ defined as:

\begin{equation}
\label{mu}
\mu = \frac{M_{HI}+M_{H_{2}}}{M_{HI}+M_{H_{2}}+kL_{B}},
\end{equation}

where $M_{H{I}}$ is the mass in atomic hydrogen, $M_{H{2}}$ is the mass in molecular hydrogen, $L_{B}$
 is the blue luminosity, and $k$ is the mass to luminosity ratio, here fixed to a value of $k=1.5$, following
the work of Garnett (2002). The value of $\mu$ reported in PVC, is plotted as a function of $\lambda$,
estimated through equation (5), and type, in a logarithmic scale presented in figure 5, where the solid line is a fit
 to the data corresponding to a relation between $\mu$ and $\lambda$ of the form:

\begin{equation}
\label{logmu}
log(\mu) =0.384(\pm 0.159) log(\lambda) + 0.082(\pm 0.097),
\end{equation}

with a correlation coefficient of $0.562$, again, equation (11) gives also the $1 - \sigma$ confidence intervals
on the fitted parameters, given our uncertainties in $\lambda$. The average value of the dispersion shown by the data in figure 6
is of 0.167 dex. The clear trend, in the expected sense given our hypothesis, is as 
clear as that obtained with the rotational velocity, which presents a correlation coefficient of $-0.605$,  
slightly better than what is observed against the luminosity, with a correlation coefficient of $-0.477$. In comparison 
with type, the trend is similar but in this case the allocation of $\lambda$ is completely objective and 
quantitative, which allow us to calculate empirical correlations as equation (11), that would be senseless to try 
using type, given its qualitative nature. 

Thinking of $\lambda$ as the main physical determinant of galactic type,
the large gas fractions and low metallicities found for both late type spirals and high $\lambda$ systems in the sample
we treat, are consistent with the expectations of section (3). High $\lambda$ disks have long star formation
timescales, and hence appear today as low metallicity, high gas fraction systems. One would hence also expect 
these galaxies to show high present star formation rates, as low $\lambda$ systems would have exhausted their gas fuel
in the remote past. We note that Berta et al. (2008) recently adopted a slightly modified  
and somewhat refined, and higher accuracy, version of equation (5) to determine $\lambda$ for a large sample (52,000) of 
SDSS galaxies, and strongly confirm the relation between this physical parameter
and galactic type. In the particular case of the recent star formation, which they accurately trace through detailed
spectral synthesis modeling, they find a positive correlation with $\lambda$, which is what would be
expected from the preceding argument. The work of Berta et al. (2008) offers independent support for both the 
high gas mass fractions and the low metallicities of high $\lambda$ galaxies we found, after the critical 
dependance on total mass has been considered.
Further, in that work the authors strongly support the identification of $\lambda$ and mass 
as the parameters determining the Hubble sequence.

\begin{figure}[htb]
\begin{tabular}{l}
\includegraphics[width=\columnwidth]{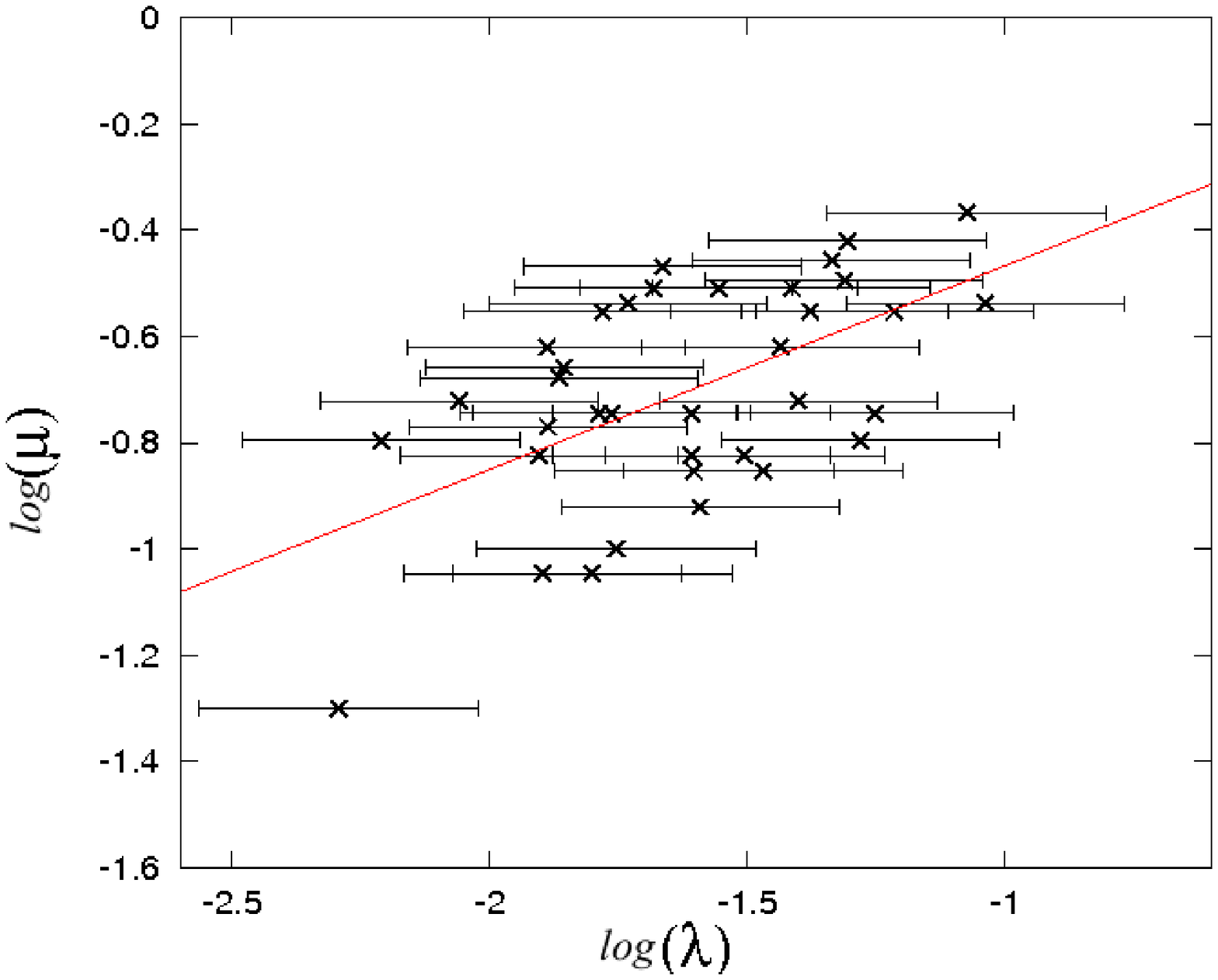}\\
\includegraphics[width=\columnwidth]{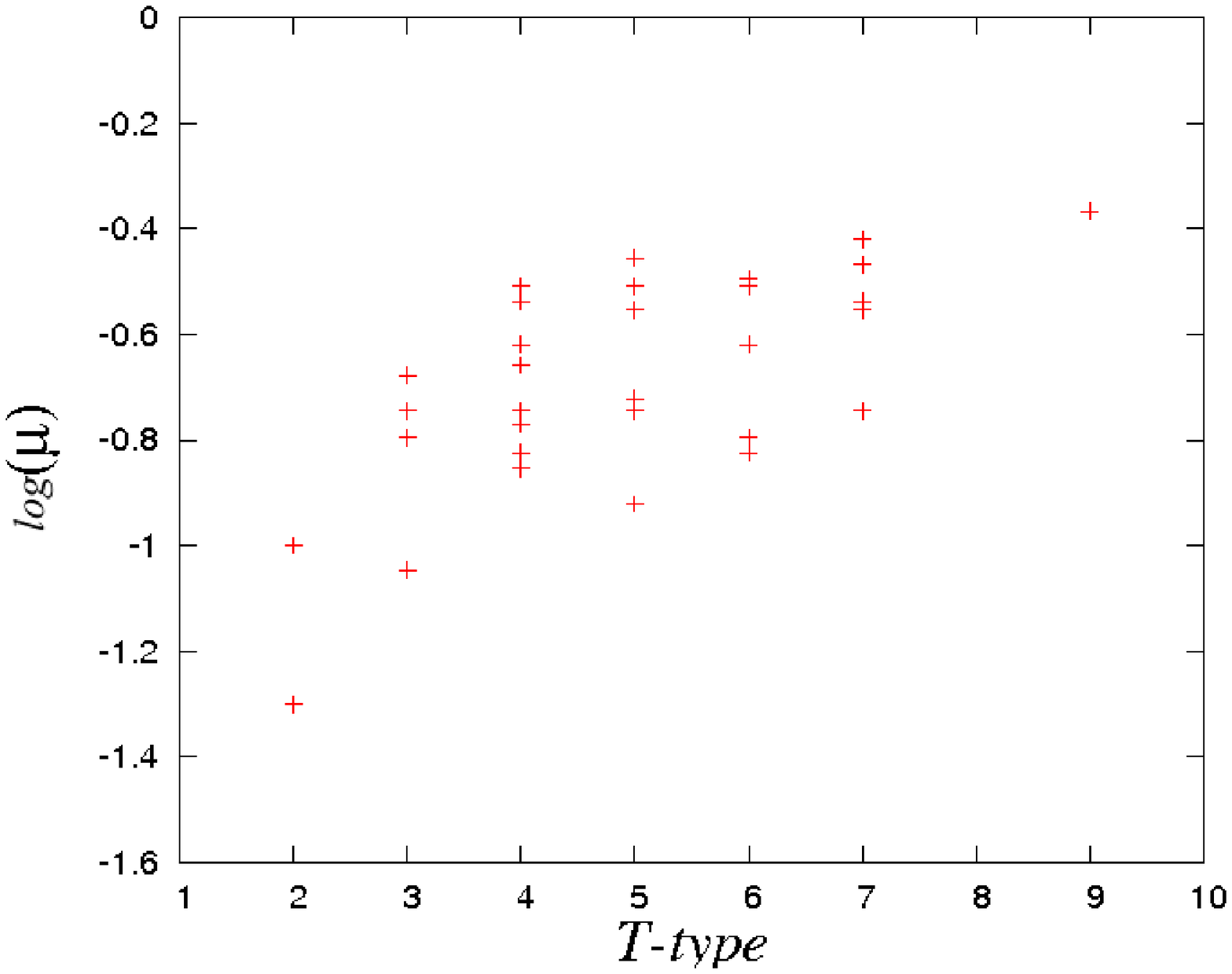}
\end{tabular}
\label{histograms}
\caption[ ]{Logarithmic plot of the gas fraction of the 36 galaxies from the PVC sample, as a function of
 $\lambda$ (top panel) and type (bottom panel). The solid line is the best fit to the data.}
\end{figure}

\section{Conclusions}

We use a simple dynamical model for spiral galaxies to inferring the value of $\lambda$ for a
sample of observed disk galaxies, for which detailed chemical and gas mass studies exist.

We confirm general results coming from theoretical studies of chemical galactic evolution; systems with low
$\lambda$ spin parameter present typically higher metallicities and lower gas mass fractions than 
systems with large $\lambda$. In low mass disk galaxies, the abundance gradients in $dex$ $kpc^{-1}$ are more 
pronounced for low $\lambda$ systems. Asides from the trends found, we present numerical relations between
the variables involved which could be used to calibrate numerical studies of galactic formation. 

The trends here discussed together with correlations between $\lambda$ and other physical observable parameters
such as the disk thickness, the bulge to disk ratio, colour and colour gradients, as stressed by several authors
(van der Kruit 1987, Dalcanton, Spegel \& Summers 1997, van den Bosch 1998,
Hernandez \& Cervantes-Sodi 2006, Hernandez, Park, Cervantes-Sodi \& Choi 2007), reinforce the idea of the 
use of this physical parameter as an objective and quantitative tool for galaxy classification.

It appears clear that once the total mass of a galaxy has been fixed, and the integral properties 
hence determined (e.g. through the Tully-Fisher relation), it is $\lambda$ what then establishes the 
internal structure, disk scale lengths, star formation efficiencies and other derived quantities.

\section{Acknowledgments}
The authors acknowledge the constructive criticism of an anonymous referee in helping to reach a more balanced
final version. The work of B. Cervantes-Sodi was supported by a CONACYT scholarship. The work of X. Hernandez was 
partially supported by DGAPA-UNAM grant no IN114107.

\end{document}